\documentclass[acmtog,nonacm]{acmart}
\makeatletter
\let\hyxmp@parse@acmart\relax
\makeatother
\hypersetup{
  pdfpublisher={},
  pdfpublication={},
  pdfpubtype={},
  pdfvolumenum={},
  pdfissuenum={},
  pdfdoi={},
  pdfcontactemail={ds867@njit.edu; przem@njit.edu}
}
\graphicspath{{figures/}}
\title[The Shape of Speech]{The Shape of Speech: A Geometric Measure of Coarticulation for Speech-Driven 3D Facial Animation}
\author{Danzel Serrano}
\email{ds867@njit.edu}
\affiliation{%
  \institution{New Jersey Institute of Technology}
  \country{USA}
}
\author{Przemyslaw Musialski}
\email{przem@njit.edu}
\affiliation{%
  \institution{New Jersey Institute of Technology}
  \country{USA}
}
\renewcommand{\shortauthors}{Serrano and Musialski}
\begin{document}

\begin{abstract}
Speech-driven 3D facial animation can reproduce recognizable mouth poses. However, it can
simplify the motion between them, and that motion carries \emph{coarticulation}, the way the
sounds around each sound shape its articulation. We introduce a geometric measure of this
trajectory shaping: lip-path length compared with the shortest route through the vowel,
consonant and vowel positions of a speech segment. In contrast to the endpoint chord, this
consonant-aware route accounts for obligatory transit and avoids degeneracy, while preserving
invariance to uniform motion gain. The measure needs only a forced alignment, so it applies
where no ground truth exists. We demonstrate it on four state-of-the-art methods, one per
architectural family, real-time and offline. All four trace flatter lip trajectories than
captured speech. Against frame-rate-matched ground truth, DiffPoseTalk, ARTalk and FaceFormer
show clear deficits, equivalent on this measure to removing 15--60\% of real speech's fast
articulatory component. CodeTalker is marginal on the primary measure and clear on a companion
measure. A pre-registered study with 97 viewers and 3,523 judgments underpins the measured direction: controlled
damping of real motion lowers the score and is penalized, whereas exaggeration shows no
detected penalty over the tested range. Viewers also prefer real speech in $73.4\%$ of
sentence comparisons and, in the aggregate, on single words. Together, the measure, its calibration and
the study identify a perceptually relevant loss of trajectory shaping and a concrete target
for improving synthesized articulation.

\end{abstract}



\begin{teaserfigure}
  \centering
  \includegraphics[width=0.94\linewidth,trim=0bp 60bp 0bp 0bp,clip]{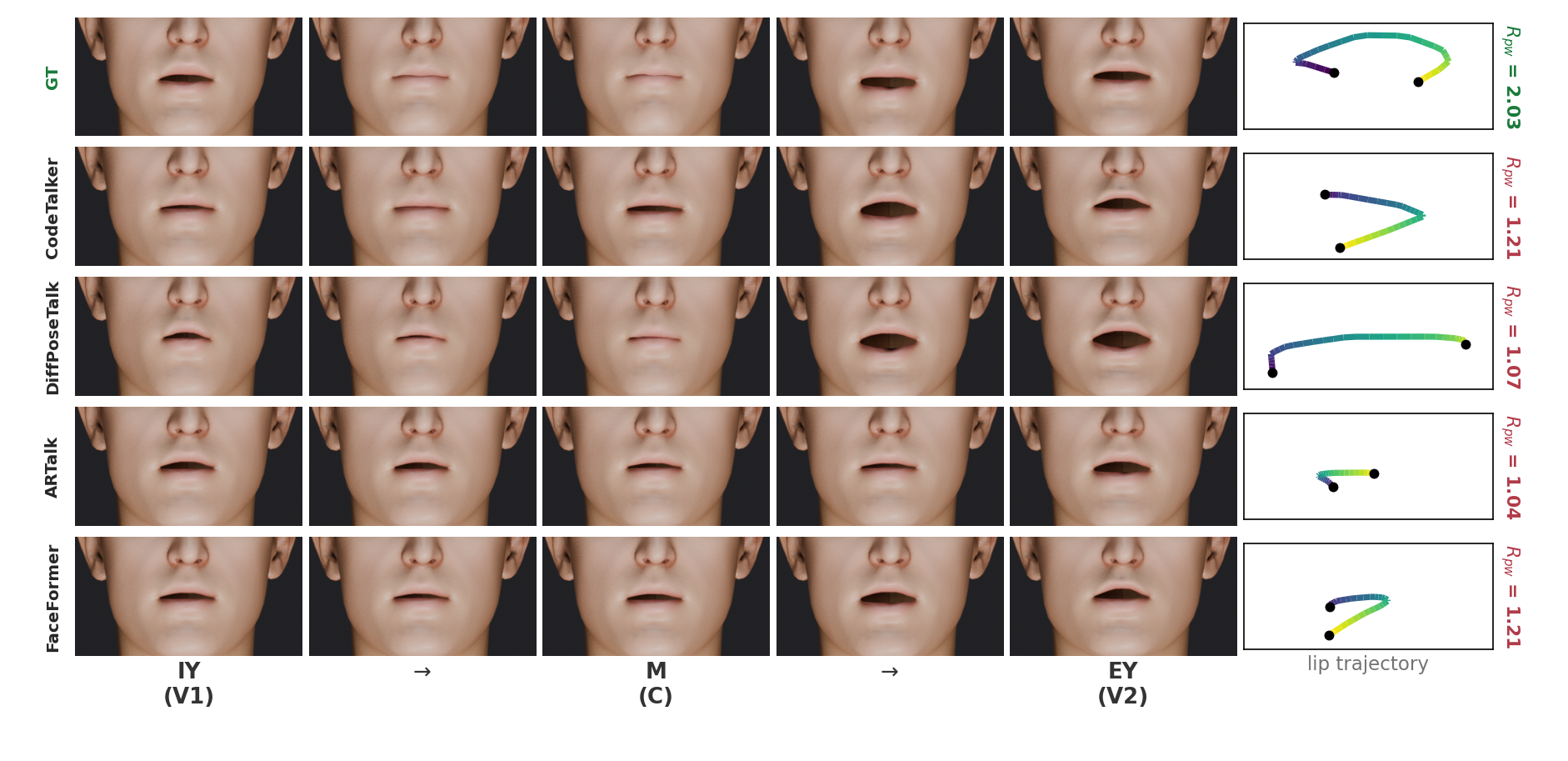}
  \caption{Similar speech poses hide flatter motion paths. Ground truth and four speech-driven animation methods on /IY-M-EY/, cropped to the lower face. Each row shows five frames from $V_1\rightarrow C\rightarrow V_2$, the lip-region trajectory over that interval, and its score $R_{pw}$ (Section~\ref{sec:metric}): 2.03 for ground truth against 1.04--1.21 for the methods. The illustration uses the 60\,fps ground-truth timeline with upsampled method outputs, while statistical comparisons use frame-rate-matched ground truth.}
  \label{fig:teaser}
\end{teaserfigure}

\maketitle

\section{Introduction}
A talking face can reach recognizable mouth shapes in time with speech. However, it can still
move unnaturally. The paths between those shapes carry part of the articulation. For instance,
a /m/ between spread vowels closes the lips along a path shaped by its neighbors.
Figure~\ref{fig:teaser} shows this coarticulation in a captured /IY-M-EY/ sequence and four
speech-driven 3D facial animation (SD3DFA) methods. Their trajectories reveal differences in
movement that individual poses do not convey.

Lip vertex error measures how close predicted geometry is to a reference. Lip-sync measures
whether mouth motion matches the audio's timing. However, neither is designed to isolate the
shape of the path between speech targets. To address this, we introduce a measure of that
shape: the lip path compared with the shortest route through the lip positions at the vowel,
consonant and vowel of a speech segment. The ratio discounts uniform motion gain, so increased
amplitude alone cannot improve it, and it needs only a forced alignment.

We demonstrate the measure on four state-of-the-art methods, one per architectural family,
real-time and offline. The evaluated animations flatten these paths systematically. Relative
to captured speech at each method's frame rate, DiffPoseTalk~\cite{sun2024diffposetalk}, ARTalk~\cite{artalk2025} and FaceFormer~\cite{fan2022faceformer} have
clear deficits in residual lip-path shaping. CodeTalker's~\cite{xing2023codetalker} deficit is marginal on our primary
measure and clear on a companion statistic. Calibrating the deficits against controlled
damping of real motion makes their magnitude interpretable. In a pre-registered user study
with 97 viewers and 3,523 judgments, viewers penalize damping along the same measured
direction. This combination identifies an animation deficit
that matters for developing more natural speech motion.

Our practical motivation is to improve synthesized articulation. A measurable deficit provides
a target for training and motion control: preserve the trajectory shaping found in real
speech. Our contributions are three-fold:
\begin{itemize}
  \item A \textbf{geometric measure and its unit}: we introduce the piecewise ratio $R_{pw}$,
        which scores lip-path shape against the route through vowel, consonant and vowel. It
        is invariant to motion gain. It needs only a forced alignment, so it applies where no
        ground truth exists. Calibration against controlled damping of real speech yields the
        $\beta$-equivalent. It states a deficit as the fraction of real speech's fast
        articulatory component retained. It also lets a target value set how strongly an
        animation articulates (Sections~\ref{sec:metric} and~\ref{sec:dial}).
  \item A \textbf{demonstration on four state-of-the-art methods} (real-time and offline):
        all four trace flatter lip paths than captured speech. For two methods, the deficit
        is equivalent to removing roughly half of real speech's fast articulatory component
        (Section~\ref{sec:eval}).
  \item A \textbf{pre-registered perceptual characterization}: across 97 viewers and 3,523
        judgments, we show that viewers penalize articulation damped below real speech, but
        not exaggerated articulation. They prefer real speech in $73.4\%$ of sentence
        comparisons, and on single words they prefer it to the method output by almost two
        to one.
\end{itemize}

\section{Related Work}\label{sec:related}
\paragraph{Articulation in facial animation} Coarticulation has long been central to visual
speech synthesis. Cohen and Massaro~\cite{cohen1993modeling} use dominance functions with
overlapping, phoneme-specific weights to let neighboring segments shape each articulatory
target. Subsequent systems inherited coarticulation from recorded triphones~\cite{bregler1997video},
learned phoneme targets and their spread~\cite{ezzat2002trainable}, separated coarticulation
from expression~\cite{deng2006expressive}, and replaced static visemes with dynamic
ones~\cite{taylor2012dynamic}. JALI~\cite{edwards2016jali} exposes independent jaw and lip
controls for articulation strength. VisemeNet~\cite{zhou2018visemenet} predicts these
animator-facing curves from speech. Editable factors also appear in learned systems that
separate emotion from content~\cite{peng2023emotalk} or articulation from
prosody~\cite{serrano2025disentangled}. What is still missing is a way to \emph{measure} the
trajectory shaping that this control acts on.

Recent SD3DFA methods build on learned audio-to-mesh and phoneme-to-animation
mappings~\cite{karras2017audio,taylor2017deep}, VOCASET and VOCA on FLAME
topology~\cite{cudeiro2019voca,li2017flame}, and MeshTalk's cross-modality
disentanglement~\cite{richard2021meshtalk}. We evaluate four representatives with different
motion representations. FaceFormer~\cite{fan2022faceformer} predicts continuous vertex offsets
with an autoregressive transformer. Related continuous models add
personalization~\cite{thambiraja2023imitator} or self-supervised
consistency~\cite{peng2023selftalk}. CodeTalker~\cite{xing2023codetalker} introduces a discrete
motion codebook to address over-smoothed regression outputs.
DiffPoseTalk~\cite{sun2024diffposetalk} uses diffusion over motion and pose, as does
FaceDiffuser~\cite{stan2023facediffuser}. ARTalk~\cite{artalk2025} predicts multi-scale
residual motion tokens for real-time animation. Other approaches relax the fixed-topology
assumption~\cite{nocentini2024scantalk} or train across heterogeneous
annotations~\cite{fan2024unitalker}.

\begin{figure}[t]
  \centering
  \includegraphics[width=\columnwidth]{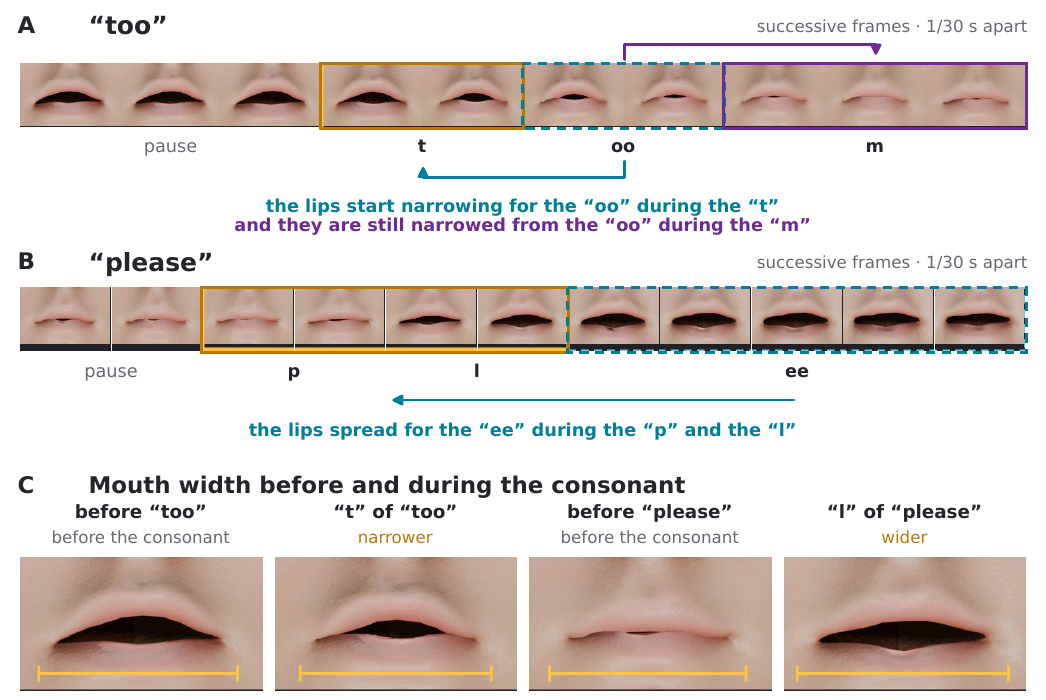}
  \caption{The consonant takes the shape of the vowel around it. \textbf{(A)} ``too'', with anticipation on the ``t'' and carryover on the following ``m''. \textbf{(B)} ``please'', with the lips spreading for the ``ee'' during the ``p'' and ``l''. Successive frames are 1/30\,s apart. \textbf{(C)} Mouth width before and during the consonant, at equal scale: narrower during the ``t'', wider during the ``l''. See the supplementary video for an animated explanation of coarticulation.}
  \label{fig:coarticulation}
\end{figure}

\begin{figure*}[!t]
  \centering
  \begin{minipage}[t]{0.49\textwidth}\centering
    \includegraphics[width=\linewidth]{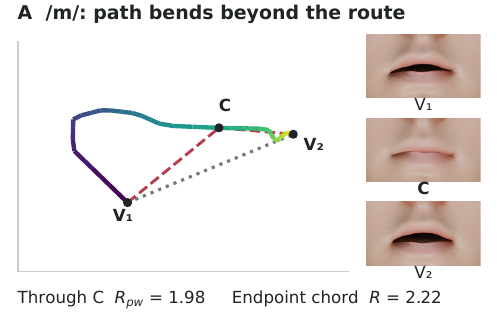}
  \end{minipage}\hfill
  \begin{minipage}[t]{0.49\textwidth}\centering
    \includegraphics[width=\linewidth]{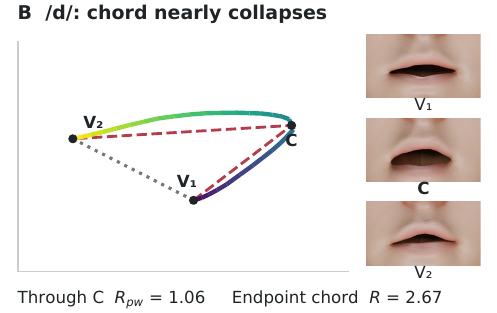}
  \end{minipage}
  \caption{The consonant changes the baseline. Real lip trajectories with illustrative V$_1$, C and V$_2$ mouth poses. Time runs along each colored path. Gray dotted lines join only the endpoints, while red dashed lines go through C. The route-based $R_{pw}$ distinguishes the bent /m/ path (1.98) from /d/ (1.06). However, endpoint $R$ reverses them (2.22 versus 2.67).}
  \label{fig:metric}
\end{figure*}

\paragraph{Evaluation of speech animation} Lip-sync evaluation commonly uses audio-visual
networks such as SyncNet~\cite{chung2016out} and the confidence and distance measures
associated with Wav2Lip~\cite{prajwal2020lip}. Lip vertex error aggregates pointwise geometric
differences. CodeTalker's upper-face dynamics deviation~\cite{xing2023codetalker} compares
temporal standard deviations and therefore measures motion amplitude. However, these measures
address properties other than residual lip-path shape. Nocentini et
al.~\cite{nocentini2025beyond} likewise argue for evaluation beyond standard measures to
capture perceptually relevant differences. Haque et al.~\cite{haque2025wild} report that common objective scores disagree with
perceptual ratings and that method rankings do not transfer across datasets. Lee et
al.~\cite{lee2025perceptual} define perceptual criteria for 3D lip sync and matching
metrics. Our path ratio needs only phone timings from forced
alignment, not matched ground truth. Its validation here is comparative, using captured speech
on the same tokens.

\paragraph{Perception of animated motion} Sensitivity to motion differences and impressions
of animated humans depend on rendering~\cite{hodgins1998perception,mcdonnell2012render}. Hyde et
al.~\cite{hyde2013perceptual} measured the perceptual effects of damped and exaggerated
facial motion.
Reitsma and Pollard~\cite{reitsma2003perceptual} calibrated an animation measure against
perceptual thresholds for ballistic motion. Evaluation methods for virtual humans have since
been surveyed~\cite{rekik2024survey}. We test the perceptual consequences of a controlled
motion change, preferences for synthesized speech, and whether those preferences scale with
our measured deficit (Section~\ref{sec:perceptual}).

\paragraph{Coarticulation in speech science} The degree-of-articulatory-constraint
account~\cite{recasens1997model} predicts that a sound resists contextual reshaping on
articulators it uses for its own constriction. The glottal /hh/ of \emph{ahead} imposes no oral
constriction, so it should add little residual shaping to the lip path. We test this
prediction in Section~\ref{sec:metric}. Coarticulation also attenuates across word and prosodic
boundaries~\cite{cho2004prosodic,cho2008transboundary,hardcastle1999coarticulation}. We examine
word scope in Section~\ref{sec:eval}. These accounts guide our measurement on full 3D
lip-region trajectories, rather than acoustics or the sparse points tracked by articulography.

\section{Background: Coarticulation}\label{sec:background}

Speech sounds are not produced one at a time. The mouth moves continuously, and the movement
toward each sound is shaped by the sounds around it. The same consonant is therefore
articulated differently in every context. This context dependence is called
\emph{coarticulation}. It arises because the movements for successive sounds overlap in time.
First, the lips begin preparing an upcoming sound while still completing the current one,
which phonetics calls \emph{anticipation}. Second, a completed sound leaves its trace on the
motion that follows, called \emph{carryover}.

Figure~\ref{fig:coarticulation} shows both in captured speech. In ``too'', the lips are
already narrowing for the ``oo'' while the ``t'' is still being formed (anticipation), and
they remain narrowed during the following ``m'' (carryover). In ``please'', they
spread toward the ``ee'' during the ``p'' and ``l''. Panel C measures the consequence
directly: the mouth is narrower during the ``t'' and wider during the ``l'' than before the
consonant began. In other words, the consonant is not a fixed pose inserted between vowels.
It is produced inside a continuing vowel-to-vowel movement and takes its shape from it.

Speech science studies this machinery in full, with the tongue, jaw and larynx tracked
directly. However, a talking face exposes only part of it, so for graphics we opt for a
deliberately simplified view: the motion of the visible lip region.

{\"O}hman~\cite{ohman1966coarticulation} formalized this for vowel--consonant--vowel
sequences: the consonant acts as a perturbation superimposed on an underlying vowel-to-vowel
gesture that never stops. Articulatory phonology~\cite{browman1992articulatory} generalizes
the view, describing utterances as overlapping gestures rather than concatenated segments.
Two consequences matter
for this paper. First, the lip path between speech targets is not incidental transit but
carries the structure of speech production. A path that merely visited each target in a
straight line would thus be the signature of concatenation, not of speech. Second, what
distinguishes real from concatenated motion is the shape of that path, not how far the lips
travel. The next section turns both observations into a measure.

\begin{figure}[!t]
  \centering

  \includegraphics[width=\linewidth,trim=11.5bp 0.5bp 10bp 0bp,clip]{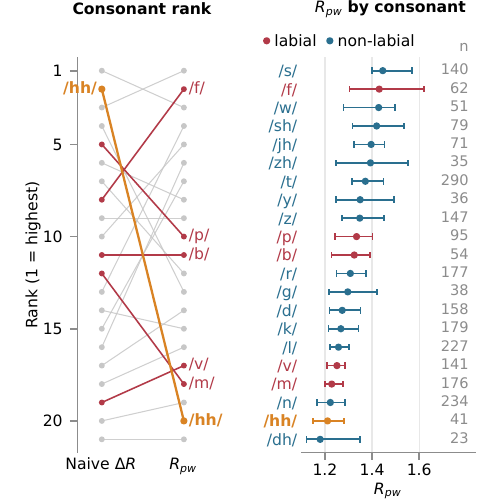}
  \caption{The route through C ranks /hh/ low. Among 21 consonants with $n\ge20$, /hh/ moves from second under the endpoint chord to twentieth under the route through C, consistent with its weak oral target. The right panel shows per-consonant estimates and speaker-clustered 95\% bootstrap intervals. Labials (red) scatter through the ordering rather than leading it.}
  \label{fig:consonants}
\end{figure}

\section{Measuring Lip-Trajectory Shaping}\label{sec:metric}
\paragraph*{Tokens and trajectories.} Phone boundaries come from the Montreal Forced
Aligner~\cite{mcauliffe2017montreal}. A \textbf{VCV token} spans
vowel$_1$--consonant--vowel$_2$. A \textbf{VV token} spans two vowels without an intervening
consonant. It supplies the baseline of the naive measure. Each token is tagged
\emph{within-word} or \emph{cross-boundary} from its alignment. We resample tokens to $T$
points in normalized time to compare different durations. Let $x(t)\in\mathbb{R}^{N\times 3}$
denote the positions of a fixed region of $N$ lip vertices relative to the subject's neutral
template. Distances use the per-vertex root-mean-square norm
$\lVert a\rVert = \big(\tfrac{1}{N}\sum_{v}\lVert a_v\rVert_2^2\big)^{1/2}$.
The token's path length is
\begin{equation}
  L \;=\; \sum_{t=1}^{T-1} \lVert x(t{+}1) - x(t)\rVert .
  \label{eq:arclen}
\end{equation}

\paragraph*{The consonant-aware route.} Let $c$ be a waypoint on the trajectory. We define
\begin{equation}
  R_{pw} \;=\; \frac{L}{\lVert x(1) - c\rVert + \lVert c - x(T)\rVert}.
  \label{eq:Rpw}
\end{equation}
The denominator is the shortest route through the waypoint. Thus, $R_{pw}>1$ measures path
length beyond it. We place $c$ at the trajectory's position at the \emph{forced-aligned
consonant center}, and $x(1),x(T)$ at the centers of the flanking vowels. Each of these three
\emph{knots} lies on the trajectory at a linguistically defined instant. In contrast to
alignment boundaries, which fall mid-transition, vowel centers make the interval run between
speech targets.

\paragraph*{The endpoint chord.} A natural measure of detour compares path length with
straight-line displacement:
\begin{equation}
  R \;=\; \frac{L}{\lVert x(T) - x(1)\rVert},
  \label{eq:R}
\end{equation}
where a straight path has $R=1$. Subtracting the VV baseline gives
\begin{equation}
  \Delta R \;=\; R_{\mathrm{vcv}} - R_{\mathrm{vv}},
  \label{eq:dR}
\end{equation}
which is intended to isolate the consonant's additional detour. However, the chord ignores the
obligatory transit through the consonant position and counts it as detour. If the flanking
vowels nearly coincide, the chord also shrinks toward zero and inflates $R$ regardless of what
the consonant does. Figure~\ref{fig:metric} illustrates the baseline problem.

The chord also leaves a gesture-magnitude confound. A larger consonant excursion increases
path length without a corresponding change in the endpoint displacement. The evaluated
methods' median per-token lip-motion amplitudes range from $0.9$ to $1.8$\,mm, a factor of
two. Uniform rescaling of an entire trajectory cancels from $R$, but changing its interior
excursion does not. The baseline must account for that excursion before measuring residual
shape.

\paragraph*{Gain invariance.} Under a uniform scaling $x\mapsto\gamma x$, path length becomes
$L\mapsto\gamma L$. The on-path knots scale with it, so the denominator of
Equation~\ref{eq:Rpw} scales identically. Note that the naive $R$ shares this invariance.
Replacing the chord corrects the baseline accounting and its degeneracy, while preserving gain
invariance. In contrast, a fixed off-path waypoint fails to scale with the motion and loses
this property (Section~\ref{sec:sensitivity}). Both ratios also use only positional
differences and are translation-invariant, so the template-relative reference frame does not
affect them.

The two examples in Figure~\ref{fig:metric} illustrate the denominator, not consonant-class
behavior. The /m/ is an upper-tail token. A clean bilabial closure and return can remain close
to the route and score low (Figure~\ref{fig:consonants}, right). Since the waypoint follows the motion,
the denominator absorbs the excursion reached at the consonant center. This excursion includes
both the consonant's gesture and its displacement by neighboring vowels. In other words,
$R_{pw}$ measures \emph{residual lip-path shaping beyond the three knots}. It does not
estimate the absolute amount of coarticulation or gesture magnitude.

\paragraph*{Temporal sampling.} Arclength depends on sampling density, so neither ratio has a
frame-rate-independent value. Resampling ground truth from 60\,fps lowers $R_{pw}$ by $0.076$
at 30\,fps and $0.101$ at 25\,fps, exceeding the method gaps in Table~\ref{tab:piecewise}. An
unmatched comparison would therefore mostly measure sampling rate. That is why every method
comparison uses ground truth resampled to the method's own frame rate (``matched-fps GT'').

\paragraph*{A phonetic check.} English /hh/ has no appreciable vocal-tract
constriction~\cite{ladefoged1996sounds} and is underspecified for oral
place~\cite{keating1988underspecification,garellek2023voicing}. In a V--/hh/--V sequence, the
lips largely follow the vowel-to-vowel transit. The constraint account
(Section~\ref{sec:related}) therefore predicts low \emph{residual shaping}, even though
/hh/ is highly susceptible to contextual coarticulation because it resists little.

We have tested this prediction on 2492 VCV tokens from all twelve VOCASET speakers, with
bootstrap intervals clustered by speaker. Under $\Delta R$, /hh/ ranks second of twenty-one
consonants (cf.~Figure~\ref{fig:consonants}). Its assimilated flanking vowels expose the chord
degeneracy. In contrast, under $R_{pw}$ the same tokens place /hh/ twentieth, consistent with
the prediction. It also ranks last of twenty-one when endpoints are taken at token boundaries.
The result thus survives the endpoint convention.

Labials do not exceed non-labials under $R_{pw}$. The class difference is $-0.041$
[$-0.071$, $-0.014$], small but reliable. The measure is therefore not simply a
bilabial-closure detector. Individual-phone rankings are unsupported, since adjacent
intervals overlap heavily and order depends on the construction (Spearman $\rho=0.46$ between
the two measures' orderings).

\section{The Deficit and Its Calibrated Magnitude}\label{sec:eval}
\paragraph*{Evaluation setup.} We evaluate CodeTalker~\cite{xing2023codetalker}, FaceFormer~\cite{fan2022faceformer}, DiffPoseTalk~\cite{sun2024diffposetalk} and ARTalk~\cite{artalk2025} on
VOCASET's~\cite{cudeiro2019voca} standard two-subject held-out split, the same split as FaceFormer and CodeTalker.
Each method runs unchanged on held-out audio and produces FLAME~\cite{li2017flame} motion (5023 vertices). FaceFormer and
CodeTalker were trained on VOCASET. DiffPoseTalk and ARTalk run zero-shot, as released. Ground
truth yields 421 VCV tokens. The common-output set contains 183 tokens in 38 utterances. All
method comparisons use this set and matched-fps ground truth. We use the template-relative lip
trajectories of Section~\ref{sec:metric}. All intervals in this section are percentile
bootstrap CIs from 2000 replicates clustered by utterance, accounting for dependent tokens
within an utterance.

\begin{figure}[!t]
  \centering
  \includegraphics[width=\columnwidth]{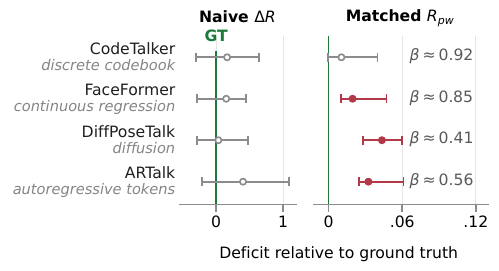}
  \caption{The route-based measure reveals flatter synthesized motion.
  Deficits relative to frame-rate-matched ground truth for four methods; positive values mean
  lower method scores. Bars are utterance-clustered 95\% bootstrap intervals, and green lines
  mark zero deficit. Left: all intervals for the endpoint-based $\Delta R$ comparison cross
  zero. Right: three $R_{pw}$ intervals clear zero; CodeTalker's crosses it. The $\beta$ labels
  express each deficit as an equivalent real-motion setting (Section~\ref{sec:dial}). The
  panels use different scales.}
  \label{fig:results}
\end{figure}

\paragraph*{Flatter paths in synthesized speech.} Three methods have clear deficits in
$R_{pw}$ relative to matched ground truth (cf.~Table~\ref{tab:piecewise},
Figure~\ref{fig:results}). CodeTalker is marginal at $+0.010$ [$-0.001$, $+0.040$]. Note that
DiffPoseTalk shows the largest paired deficit, despite having the largest lip-motion amplitude
of any arm.

A companion statistic divides the mean distance from the trajectory to that route by token
motion amplitude. Motion amplitude is the root-mean-square over lip vertices of their temporal
standard deviations. The statistic is also gain-invariant when the knots lie on the path. All
four methods show clear deficits under it, including CodeTalker at $+0.031$ [$+0.019$,
$+0.057$], with larger gaps for the other methods. The two statistics thus agree on the
direction of the deficit, with weaker primary evidence for CodeTalker.

\begin{table}[b]
  \centering
  \caption{All four methods score below ground truth, three of them clearly. $R_{pw}$ on 183 matched tokens from 38 utterances. Levels are token medians. Gaps are median paired differences against matched-fps ground truth, with utterance-clustered 95\% CIs. They are not differences of the displayed levels and are more conservative for every method. The last column gives $\beta$-equivalents (Section~\ref{sec:dial}).}
  \label{tab:piecewise}

  \setlength{\tabcolsep}{2.4pt}\small
  \begin{tabular}{llccc}
    \toprule
    arm & fps & $R_{pw}$ & GT $-$ method [95\% CI] & $\beta$-equiv [95\% CI] \\
    \midrule
    ground truth  & 30 & \textbf{1.143} & --- & $\equiv 1$ \\
    ground truth  & 25 & \textbf{1.120} & --- & $\equiv 1$ \\
    CodeTalker    & 30 & 1.107 & $+0.010$ [$-0.001$, $+0.040$] & 0.92 [0.68, 1.02] \\
    FaceFormer    & 30 & 1.085 & $+0.020$ [$+0.010$, $+0.047$] & 0.85 [0.56, 0.93] \\
    DiffPoseTalk  & 25 & 1.055 & $+0.043$ [$+0.028$, $+0.060$] & 0.41 [0.20, 0.63] \\
    ARTalk        & 25 & 1.053 & $+0.032$ [$+0.025$, $+0.061$] & 0.56 [0.22, 0.71] \\
    \bottomrule
  \end{tabular}
\end{table}

\paragraph*{Why the naive comparison fails.} All $\Delta R$ point estimates are lower than
ground truth. However, every gap interval includes zero (Figure~\ref{fig:results}, left). Beyond the
baseline failures in Section~\ref{sec:metric}, this comparison relies on only 71 VV tokens for
ground truth and 40 for the methods. The small vowel-to-vowel baseline leaves $\Delta R$
underpowered at this scale.

\subsection{Calibrating the Deficit with Controlled Articulation}\label{sec:dial}
We manipulate captured motion to express the geometric deficit in interpretable units and to
construct stimuli for the perceptual study. Temporal Gaussian smoothing ($\sigma=3$ frames at
60\,fps) separates the vertex trajectory into a slow carrier and a fast residual. Scaling the
residual by $\beta$ (i) damps articulation for $\beta<1$, (ii) reproduces ground truth exactly
at $\beta=1$, and (iii) exaggerates it for $\beta>1$. Table~\ref{tab:dial} measures this
ladder after decimation to 30\,fps.

Both gain-invariant statistics rise monotonically with $\beta$, and $R_{pw}$ at $\beta=2$ is
interval-separated from $\beta=0$. The manipulation changes trajectory shape as well as size.
Across the ladder, amplitude grows by a factor of $2.0$, path length by $2.5$, and the
companion statistic by $2.1$. Meanwhile, $R_{pw}$ rises from $1.05$ to $1.24$. Thus,
preference along this ladder cannot be attributed to shape alone. In other words, the ladder
tests the combined consequence of damping or magnifying the fast component.

\begin{table}[b]
  \centering
  \caption{Both gain-invariant statistics rise monotonically with $\beta$. Controlled articulation ladder on all 421 ground-truth tokens at 30\,fps. Entries are medians, with utterance-clustered 95\% CIs for $R_{pw}$. The $\beta=1$ row is unmodified speech in this full set. This set differs from the 183-token matched set behind the per-fps $\beta$-equivalents in Table~\ref{tab:piecewise}. Here, resid/amp is the companion statistic, amp is motion amplitude, and $L/(T{-}1)$ is path length per normalized time step in mm.}
  \label{tab:dial}

  \setlength{\tabcolsep}{3.2pt}\small
  \begin{tabular}{lccccc}
    \toprule
    $\beta$ & $R$ & $R_{pw}$ & resid/amp & amp (mm) & $L/(T{-}1)$ \\
    \midrule
    0.0 & 1.17 & 1.053 [1.042, 1.062] & 0.121 & 0.83 & 0.10 \\
    0.5 & 1.32 & 1.093 [1.082, 1.112] & 0.166 & 1.00 & 0.13 \\
    1.0 & 1.48 & 1.157 [1.127, 1.179] & 0.200 & 1.21 & 0.17 \\
    1.5 & 1.64 & 1.206 [1.172, 1.228] & 0.226 & 1.44 & 0.20 \\
    2.0 & 1.76 & 1.240 [1.200, 1.274] & 0.250 & 1.67 & 0.25 \\
    \bottomrule
  \end{tabular}
\end{table}

\paragraph*{The $\beta$-equivalent.} We invert the monotone response to find the $\beta$ at
which real speech has a target $R_{pw}$. We map the paired method deficits through per-fps
ladders on the 183 matched tokens, carrying uncertainty through an utterance-clustered
bootstrap. These ladders are separate from the 421-token ladder in Table~\ref{tab:dial}. The
resulting $\beta$-equivalents appear in Table~\ref{tab:piecewise} and
Figure~\ref{fig:results}. On this measure, DiffPoseTalk ($\beta\approx0.41$) and ARTalk
($0.56$) resemble real speech with roughly half its fast component removed. FaceFormer
corresponds to $\beta\approx0.85$. In contrast, CodeTalker ($0.92$) is not separable from full
articulation.

Note that this is an equivalence in $R_{pw}$, not a claim that the models literally low-pass
their motion. The same inversion also provides a calibration for future motion control.

\subsection{Scope and Sensitivity}\label{sec:sensitivity}
\paragraph*{Within and across words.} All four within-word $R_{pw}$ gaps clear zero
($+0.009$ to $+0.036$, CodeTalker's narrowly). Across boundaries, DiffPoseTalk, ARTalk and
FaceFormer also clear zero ($+0.052$ to $+0.080$). However, CodeTalker's interval includes
zero in this smaller stratum ($n{=}56$). Ground-truth $R_{pw}$ is similar within and across
words ($1.228$ versus $1.244$ at 60\,fps over all 421 tokens). Boundary attenuation may
therefore act chiefly on articulation magnitude rather than residual path shape.

\paragraph*{Baseline choices.} We cross (1) two endpoint conventions (token boundaries and
vowel centers), (2) three waypoints (the aligned consonant center, the token's own maximum
deviation from the chord, and a corpus-level consonant medoid), and (3) the two statistics.
Restricting the waypoint to the trajectory leaves us with $32$ method--construction--statistic
cells. Ground truth exceeds the method with an interval clear of zero in $31$. The sole
exception is CodeTalker under the primary construction in Table~\ref{tab:piecewise}. Its other
seven cells clear zero. However, method ordering reproduces in only a minority of
constructions. The stable finding is the deficit relative to real speech.

In contrast, the fixed off-path medoid loses gain invariance and reverses the
amplitude-normalized comparison, placing methods above ground truth. This uncontrolled
comparator demonstrates the consequence of letting amplitude differences enter through the
baseline.

\section{Perceptual Evidence}\label{sec:perceptual}

In this section, we put the measured direction to a pre-registered perceptual test with
97 viewers and 3,523 judgments. In every
trial, both sides show the same generic face and play the same audio. Only the motion
differs. In other words, the study asks one question three ways: whether viewers notice what
the methods lose. Table~\ref{tab:study-results} in Appendix~\ref{app:study-results} lists response counts and estimates.

\paragraph*{Pre-registered tests.} \textbf{H1} predicts a positive word-level relationship
between the $R_{pw}$ deficit and preference for real motion. \textbf{H2} predicts preference
for real articulation over both damped and exaggerated motion, requiring all four tested rungs
to exceed neutrality. \textbf{H3} tests whether sentence-level preference for real speech
exceeds $50\%$. The $\beta$ arm directly tests the manipulation used for calibration. The
words and sentences arms compare method output with matched real motion. We fixed
hypotheses, models, exclusions and sample sizes (50 words, 32 $\beta$
and 20 sentences sessions) before collection, with no interim outcome analysis.
Statistical power depends on the stimulus pool, not only on rater count.
The supplementary material includes a video and a standalone demo of the study webpage with a reduced number of trials.

\begin{figure}[t]
  \centering
  \includegraphics[width=\columnwidth,trim=0bp 0bp 0bp 20bp,clip]{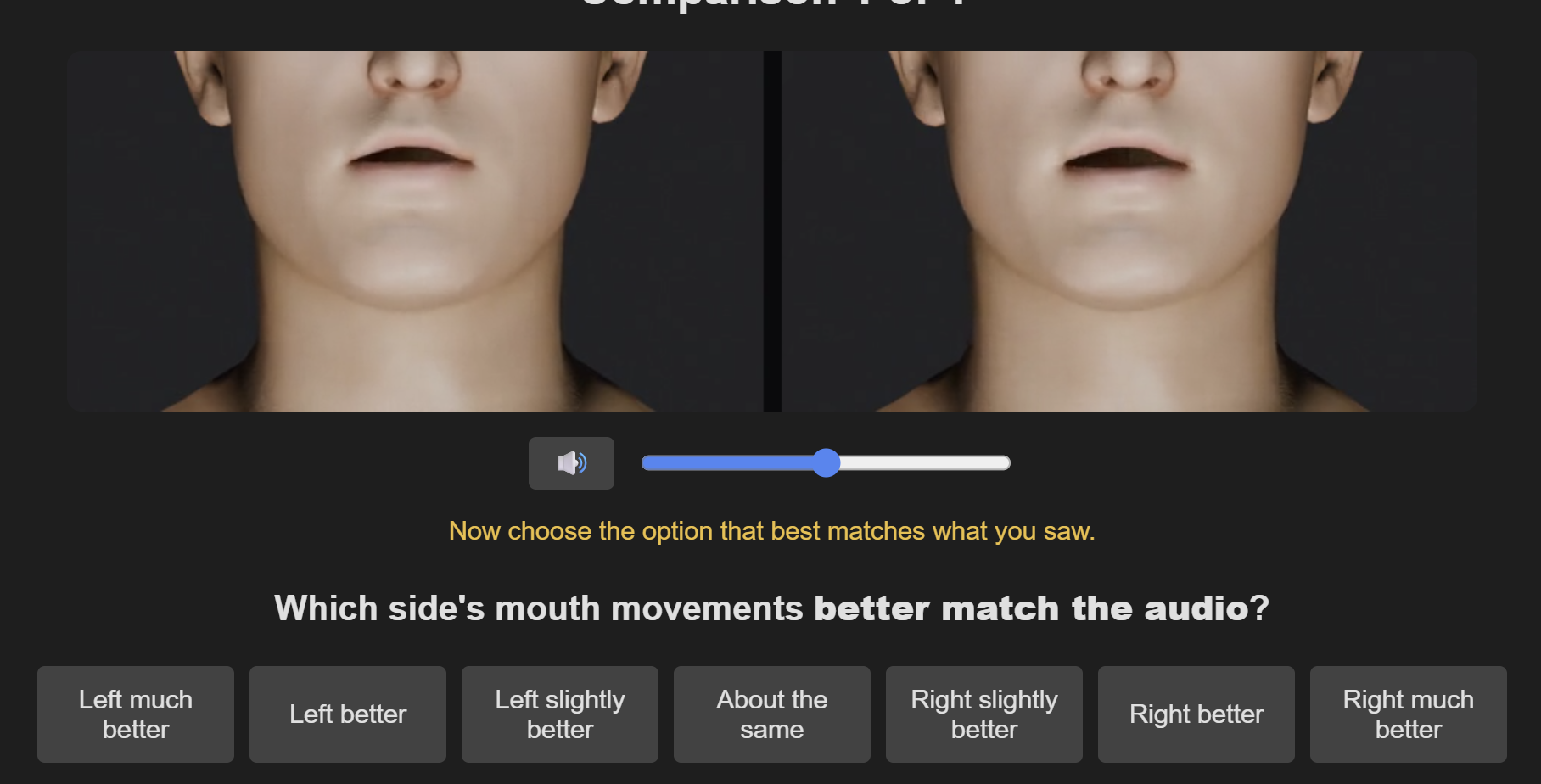}
  \caption{A trial as participants saw it. Two lower-face renders play side by side under
  the same audio, and the participant answers on a seven-point scale from ``Left much
  better'' to ``Right much better''.}
  \label{fig:trial}
\end{figure}

Participants view two animations side by side with the same audio. They rate which better
matches it on a bipolar seven-point scale, with $4$ indicating equality. We recode responses
so that higher scores favor ground truth ($g$). We crop all clips to the lower face to remove
eye and head motion from the judgment. Each session includes an unambiguous
real-against-mismatched-motion gold pair. We exclude sessions below the pre-set gold threshold
($6$ of $7$). Figure~\ref{fig:trial} shows a trial as participants saw it.

The $\beta$ arm pairs 40 tokens at $\beta\in\{0,0.5,1.5,2\}$ with their unmodified
$\beta=1$ versions. The words arm pairs all four methods with 98 within-word tokens from the
two evaluation subjects, chosen by minimum duration and common output availability. Its
per-stimulus deficit spans $[-1.45,+1.32]$ (sd $0.29$). These word-length clips last under a
second. The sentences arm pairs 19 sentences with CodeTalker, DiffPoseTalk and ARTalk. Its
stimuli predate FaceFormer's addition to the evaluation.

\paragraph*{Models.} Let $\delta_{pw}$ denote a stimulus's matched-fps
ground-truth-minus-method gap in $R_{pw}$, the per-token deficit underlying
Table~\ref{tab:piecewise}, not the vowel-baseline difference in Equation~\ref{eq:dR}. The
words model is
\[
  g \sim \delta_{pw} + \text{amp}_{\mathrm{method}} + \text{method}
      + (1\,|\,\text{participant}) + (1\,|\,\text{token}),
\]
with standardized predictors. H1 requires a positive pooled $\delta_{pw}$ coefficient with a
two-sided 95\% CI excluding zero. Including method amplitude adjusts the shape relationship
statistically for magnitude. Per-method slopes are secondary. The $\beta$ model uses the same
random effects and tests each rung's marginal preference for $\beta=1$ against neutrality.
The exaggerated rungs distinguish a preference for real articulation from a preference for
more articulation. The sentences arm compares the proportion preferring ground truth with
$50\%$ using a binomial interval.

\paragraph*{Collection and departures.} Prolific participants used desktop devices, read a
consent information sheet before starting, and were paid per session. The authors'
institutional review board determined the study exempt. After gold-pair exclusions, 54 of 71
paid words sessions, 17 of 20 sentences sessions and 26 of 32 $\beta$ sessions remain: 97
usable sessions and 3,523 comparison judgments. Three words sessions were lost to a logging
fault and appear in no count.

Two departures were decided before inspecting outcomes. First, 103 of 392 words-arm stimulus
cells had fallen below half their target. We therefore added 21 paid sessions under the
pre-registered provision for restoring coverage after exclusions. This restored 5.51 ratings
per cell against a planned 5. Second, the $\beta$ arm retained 6.50 ratings per cell against
a planned 8, since every rung already had roughly 260 ratings covering every usable
participant and token.

\begin{figure}[!t]
  \centering
  \includegraphics[width=\linewidth]{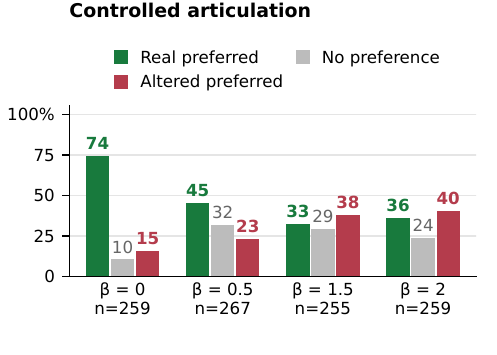}
  \caption{Damping is penalized, exaggeration is not. Each group compares real motion ($\beta=1$) with one rung of the ladder. Real motion wins clearly against $\beta=0$ and $\beta=0.5$, while against $\beta=1.5$ and $\beta=2$ the altered clip is preferred at least as often. Green favors real motion, gray is equal, and red favors the altered clip. Percentages include ties, and counts are below labels. Model estimates and intervals are reported in text.}
  \label{fig:study-beta-votes}
\end{figure}

\paragraph*{Damping real articulation has a perceptual cost.} Marginal preference for
$\beta=1$ exceeds neutrality against both damped rungs: $5.34$ (95\% CI $[5.06,5.61]$)
at $\beta=0$ and $4.31$ ($[4.03,4.58]$) at $\beta=0.5$. Real motion is preferred in
$74.1\%$ and $45.3\%$ of those trials, respectively (cf.~Figure~\ref{fig:study-beta-votes}).
However, against the exaggerated rungs, preferences of $3.86$ ($[3.58,4.14]$) at $\beta=1.5$
and $3.92$ ($[3.64,4.19]$) at $\beta=2$ do not differ detectably from neutrality. H2 asked for
all four rungs to exceed neutrality and is therefore not supported. The observed response is
asymmetric: damping is penalized, with no detected penalty for exaggeration over the tested
range. It does not support a preference peak at the real setting. This agrees with Aldeneh et
al.~\cite{aldeneh2023role}, who report a preference for over- over under-articulated lip
motion in 2D video.

\begin{figure*}[t]
  \centering
  \includegraphics[width=\textwidth]{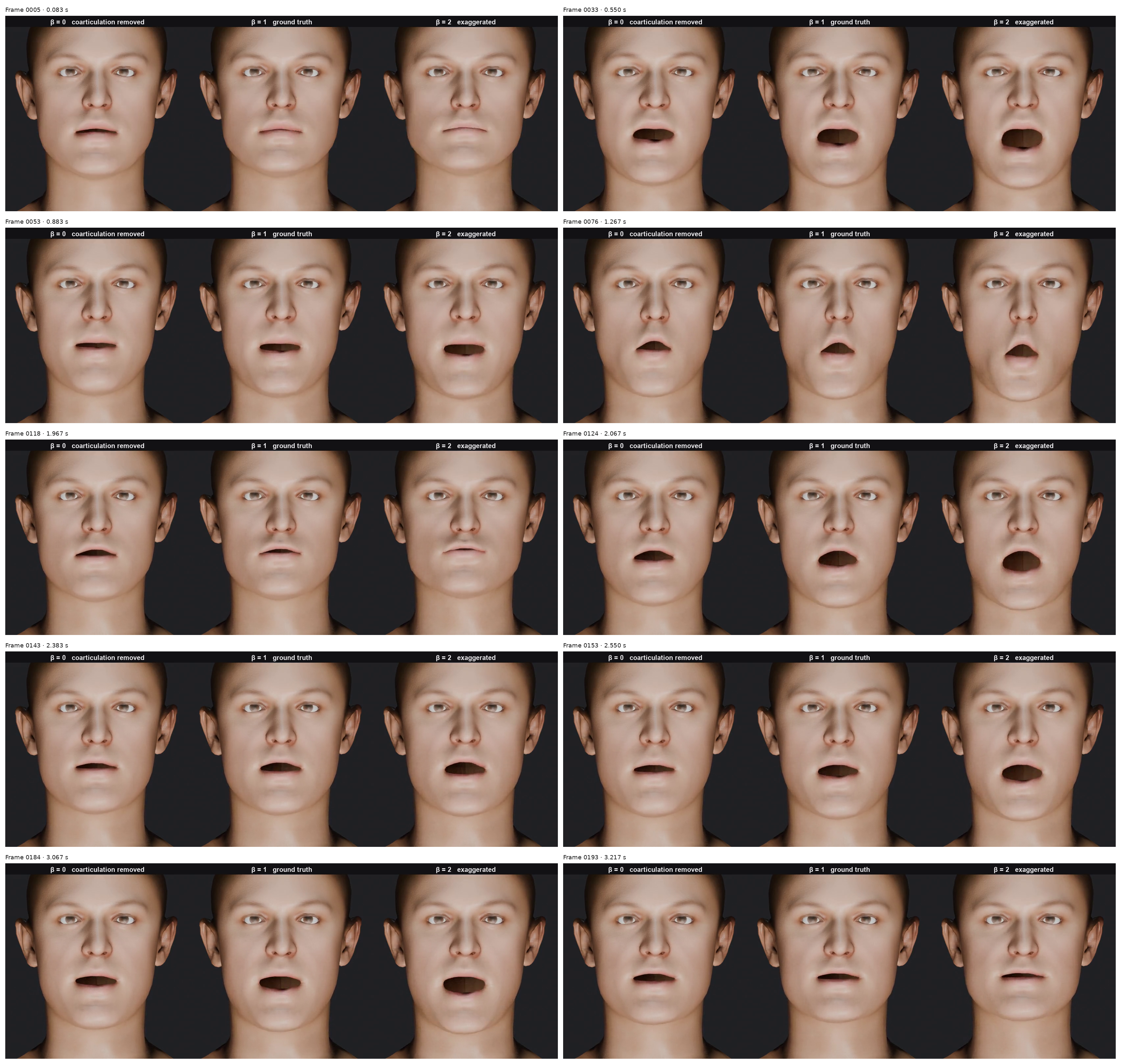}
  \caption{The articulation dial strengthens both mouth opening and closure. Controlled articulation at ten moments of captured speech. Each triptych shows $\beta=0$ (fast component removed), $\beta=1$ (unmodified speech), and $\beta=2$ (fast component doubled), from left to right. The three renderings share the same source frame, identity and view. Triptychs run in temporal order from left to right and top to bottom; labels give frame indices and times. These examples illustrate the manipulation defined in Section~\ref{sec:dial} and assessed in the $\beta$ arm of Section~\ref{sec:perceptual}.}
  \label{fig:beta-examples}
\end{figure*}

Damping lowers $R_{pw}$ in the direction of the evaluated methods' shape deficits. The
calibration places those deficits on this same tested scale, linking the geometric finding to
a controlled perceptual response. Since damping changes both shape and amplitude
(Section~\ref{sec:dial}), this link establishes the relevance of the motion change without
isolating shape as its sole cause.
Figure~\ref{fig:beta-examples} shows the visible effects of this damping and exaggeration at
ten synchronized moments.

\begin{figure*}[!t]
  \centering
  \begin{minipage}[t]{0.49\textwidth}\centering
    \includegraphics[width=\linewidth]{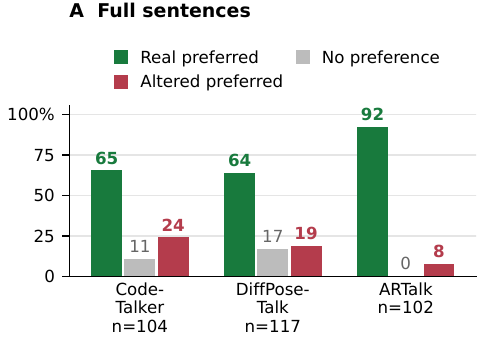}
  \end{minipage}\hfill
  \begin{minipage}[t]{0.49\textwidth}\centering
    \includegraphics[width=\linewidth]{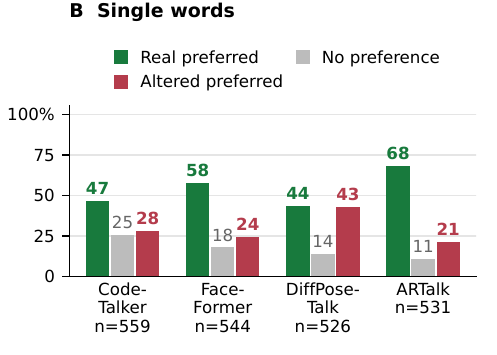}
  \end{minipage}
  \caption{Viewers prefer real speech over every method at sentence length, and more weakly on words. (A) Sentences: real motion wins 65--92\% per method, 237 of 323 trials pooled. (B) Words: margins narrow and vary by method, from parity for DiffPoseTalk to 68\% for ARTalk, with 17.0\% ties. Green favors real motion, gray is an equal rating, and red favors the method. Percentages use all trials, and counts appear beneath the method names. FaceFormer was not in the sentence arm.}
  \label{fig:study-method-votes}
\end{figure*}

\paragraph*{Viewers prefer real speech in ordinary clips.} Real motion is preferred in 237 of
323 sentence trials, $73.4\%$ (95\% CI $[68.2,78.1]$, Figure~\ref{fig:study-method-votes}).
This exceeds H3's $50\%$ threshold and replicates the pilot used to size the arm. Word-level
response shares are $53.9\%$ real motion, $17.0\%$ equal, and $29.0\%$ method motion. We read the lower word-level margin as clips too short to notice the
deficit, since the same clips reveal the larger $\beta$ manipulations reliably
(Figure~\ref{fig:study-beta-votes}). These
aggregate preferences are consistent with the geometric deficit. However, the methods differ
from real speech in other respects as well.

\paragraph*{No word-level preference predictor was established.} Across 2,160 trials from 54
sessions covering all 392 stimulus cells, the pooled $\delta_{pw}$ coefficient is $-0.033$
(SE $0.034$, 95\% CI $[-0.099,+0.033]$). H1 is not supported.
Secondary per-method slopes vary in sign, from $+0.13$ (ARTalk) to $-0.24$ (DiffPoseTalk). Viewers separate real and synthesized motion in
the aggregate. However, a per-clip mapping through $R_{pw}$ was not established. These results
do not support using the score as a proxy for individual viewer judgments during method
development.

The sensitivity analysis includes excluded sessions and changes no conclusion
(H1 $-0.031$ $[-0.092,+0.030]$, H3 $71.8\%$ $[67.0,76.3]$).

\section{Implications and Limitations}\label{sec:discussion}
The evaluated animations lose residual lip-path shaping. The calibration gives that loss a
meaningful scale. The controlled study supplies the complementary result: suppressing real
articulation moves the measure in the same direction and carries a perceptual cost. Together,
these findings make trajectory shaping a concrete target for improving SD3DFA.

\paragraph*{What the measure leaves out.} Recall that our validation is comparative. Absolute
coarticulation estimates would require a magnitude-sensitive companion measure. As a ratio of
lengths on a time-normalized trajectory, $R_{pw}$ also ignores duration and the timing of
motion between the same knots. Paths with different anticipatory or carryover timing can
receive the same score. Prosody is therefore mostly factored out by
construction, since it speaks through timing, duration and prominence, the channels the
ratio discounts. Tongue motion and other articulation not expressed by the lip region
are unobserved. This includes much of the lingual evidence behind the constraint
account~\cite{recasens1997model}. The consonant analysis supports the independently predicted
/hh/ reversal, not a complete ranking of phones.

Placing a waypoint at each interior phone would extend the construction to longer sequences.
Since every waypoint lengthens the baseline route, such comparisons would need to hold
sequence length fixed. A timing-sensitive companion could measure when the trajectory reaches
each waypoint and distinguish anticipatory from carryover effects.

\paragraph*{One face per trial, by design.} Each trial shows the same face on both sides:
one generic skin texture on the face shape of the token's speaker, one of the two held-out
speakers. Neither is a photorealistic depiction of a person. Any face still sits somewhere
on the axes of gender, age, and ethnicity, so within a trial we hold it fixed and vary
nothing but motion. Note that
this is the isolation step of a measurement, not an oversight. A motion deficit must first
be established with everything but motion held fixed. Demographic variation in the stimulus
would entangle preference with reactions to a particular appearance, which are known to
depend on rendering~\cite{hodgins1998perception,mcdonnell2012render}. The measure itself is
identity-free by construction (template-relative offsets, cf. Section~\ref{sec:metric}) and
needs no reference, so it runs unchanged on any face and any speaker. In other words, we
establish the deficit on the motion speakers share. Whether its size or visibility varies
across face morphologies, genders, or speaker groups is a direct follow-up this instrument
enables, not a question this paper answers.

\paragraph*{One corpus, one language.} All motion comes from VOCASET: controlled captures
of read English on FLAME topology. The method comparison uses two held-out speakers and 183
matched tokens, and the ground-truth analyses cover all twelve speakers. We evaluate on one
corpus for two reasons. First, VOCASET is the standard test bed of this field. FaceFormer and
CodeTalker train and report on it, and we run DiffPoseTalk and ARTalk on it zero-shot, as
released. The deficit therefore cannot be reduced to domain transfer, since it also appears
in the two methods trained on this corpus. Second, a single capture pipeline holds the
reference constant across all arms, while mixing corpora would mix method differences with
capture differences. Note that VOCASET is
our operational ground truth, not the truth itself. Capture, registration and fitting can
smooth motion, and any such smoothing removes fast motion from the reference. If anything,
the measured deficits therefore understate the gap to real human articulation. Coarticulation
patterns are also language-specific~\cite{recasens1997model}, so our numerical values
describe read English on this setup and must be validated across corpora, capture setups and
speaking styles. In contrast, the construction itself transfers: it requires only a fixed lip
region and a forced alignment. A natural next step is therefore to confirm our findings on
further corpora, languages and capture setups.

The score also isolates only part of the discrepancy between generated and real speech.
Matching $R_{pw}$ exactly moves the lips by roughly $8\%$ of the distance between a method and
ground truth. The word-level prediction test and the absence of a detected penalty for
exaggeration further limit a direct perceptual interpretation of an individual score.

\paragraph*{Future work: improving synthesis.} A first route is an auxiliary training
objective that matches generated and reference trajectory-shaping scores alongside existing
reconstruction and temporal objectives. Its target would be reference-like articulation, not
unrestricted score maximization. A second route is calibrated postprocessing: adjust generated
articulation toward the reference range and use the score to measure and control the
correction. Neither intervention has been evaluated here. The decisive experiment would
compare a baseline system and its modified version on held-out speech, measuring both the
geometric deficit and sentence-level viewer preference. Vertex-error, lip-sync and dynamics
measures would remain complementary checks on the resulting animation.

\section{Conclusion}
We have introduced a geometric measure of coarticulation for speech-driven 3D facial
animation: a consonant-aware path ratio that scores the shape of the lip path, not its size.
Applied to four state-of-the-art methods, it shows that they flatten real lip
trajectories. Calibration against controlled articulation
expresses it in units of real speech's fast motion. Three methods show clear primary deficits,
while CodeTalker is marginal on the primary statistic and clear on the companion. Viewers
penalize damping along the same measured direction, connecting the geometric finding to
perceptually relevant motion. Reference-matching training and calibrated motion correction
are concrete next steps, to be tested through geometric change and sentence-level preference
on held-out speech.

\balance
\bibliographystyle{ACM-Reference-Format}
\bibliography{references_arxiv}

\clearpage
\nobalance
\onecolumn
\appendix
\section{Perceptual Study Results}\label{app:study-results}

\begin{table}[ht]
  \centering
  \caption{Perceptual study results after the pre-registered exclusions: 97 usable sessions and 3,523 judgments.
  Response cells give counts (percentages of all trials, including ties); pooled rows summarize the rows above and add no trials.
  ``Other'' means the altered clip in the controlled-articulation arm and the method output in the sentence and word arms.
  The estimates below repeat the reported analyses: $g$ is the seven-point preference score, with $4$ neutral and higher values favoring real motion;
  the word-level coefficient uses standardized predictors. The sentence interval is binomial.
  FaceFormer was not included in the sentence arm.}
  \label{tab:study-results}
  \small
  \setlength{\tabcolsep}{5pt}
  \begin{tabular*}{0.8\textwidth}{@{\extracolsep{\fill}}llrrrr@{}}
    \toprule
    Arm & Comparison & Trials & Real preferred & Equal & Other preferred \\
    \midrule
    Controlled articulation & $\beta=0$ & 259 & 192 (74.1) & 27 (10.4) & 40 (15.4) \\
    Controlled articulation & $\beta=0.5$ & 267 & 121 (45.3) & 85 (31.8) & 61 (22.8) \\
    Controlled articulation & $\beta=1.5$ & 255 & 83 (32.5) & 75 (29.4) & 97 (38.0) \\
    Controlled articulation & $\beta=2$ & 259 & 93 (35.9) & 62 (23.9) & 104 (40.2) \\
    \addlinespace
    Sentences & CodeTalker & 104 & 68 (65.4) & 11 (10.6) & 25 (24.0) \\
    Sentences & DiffPoseTalk & 117 & 75 (64.1) & 20 (17.1) & 22 (18.8) \\
    Sentences & ARTalk & 102 & 94 (92.2) & 0 (0.0) & 8 (7.8) \\
    Sentences & Pooled & 323 & 237 (73.4) & 31 (9.6) & 55 (17.0) \\
    \addlinespace
    Words & CodeTalker & 559 & 261 (46.7) & 142 (25.4) & 156 (27.9) \\
    Words & FaceFormer & 544 & 314 (57.7) & 97 (17.8) & 133 (24.4) \\
    Words & DiffPoseTalk & 526 & 229 (43.5) & 73 (13.9) & 224 (42.6) \\
    Words & ARTalk & 531 & 361 (68.0) & 56 (10.5) & 114 (21.5) \\
    Words & Pooled & 2160 & 1165 (53.9) & 368 (17.0) & 627 (29.0) \\
    \bottomrule
  \end{tabular*}

  \medskip
  \begin{tabular*}{0.8\textwidth}{@{\extracolsep{\fill}}lllrrr@{}}
    \toprule
    Test & Comparison & Quantity & Estimate & 95\% CI lower & 95\% CI upper \\
    \midrule
    H1 & Words, pooled & $\delta_{pw}$ coefficient & $-0.033$ & $-0.099$ & $+0.033$ \\
    H2 & $\beta=0$ vs.\ $\beta=1$ & Marginal $g$ & 5.34 & 5.06 & 5.61 \\
    H2 & $\beta=0.5$ vs.\ $\beta=1$ & Marginal $g$ & 4.31 & 4.03 & 4.58 \\
    H2 & $\beta=1.5$ vs.\ $\beta=1$ & Marginal $g$ & 3.86 & 3.58 & 4.14 \\
    H2 & $\beta=2$ vs.\ $\beta=1$ & Marginal $g$ & 3.92 & 3.64 & 4.19 \\
    H3 & Sentences, pooled & Real preferred (\%) & 73.4 & 68.2 & 78.1 \\
    \bottomrule
  \end{tabular*}
\end{table}

\end{document}